\documentclass[aps,prc,preprint,tightenlines,groupedaddress,
nofootinbib,showpacs,preprintnumbers,amsmath,amssymb,floatfix,
superscriptaddress]{revtex4}
\usepackage{graphicx,psfrag,amsmath,amssymb,amsfonts,bbm}
\usepackage[latin1]{inputenc}

\newcommand{\ele}[3]{\langle #1 \mid #2\mid #3\rangle}
\newcommand{\be}{\begin{equation}}
\newcommand{\ee}{\end{equation}}
\newcommand{\ket}[1]{|#1\rangle}

\newcommand{\ba}{\begin{eqnarray}}
\newcommand{\ea}{\end{eqnarray}}
\newcommand{\bary}{\begin{array}}
\newcommand{\ear}{\end{array}}

\begin{document}

\title{A Semiclassical Model for Molecular Localization in Ammonia}

\author{Sayan Chakraborti}
\affiliation{Tata Institute of Fundamental Research (TIFR), Mumbai, India}

\begin{abstract}
The pedagogic two stste system of the ammonia molecule is used to illustrate
the phenomenon of environment induced molecular localization in pyramidal
molecules. A semiclassical model is used to describe a gas of pyramidal
molecules interacting via hard ball collisions. This modifies the tunnelling
dynamics between the classical equilibrium configurations of an isolated
molecule. For sufficiently high pressures, the model explains molecular
localization in these classical configurations. The decrease in the inversion
line frequency of ammonia, noted upon increase in pressure, is also studied.
\end{abstract}


\maketitle

\section{Introduction}
\label{intro}

An isolated ammonia molecule would periodically switch between its classical
equilibrium states \cite{Feynman}. Its stationary Eigenstates are delocalized
symmetric and antisymmetric ones \cite{Cohen}. The energy difference between
these states account for the absorption frequency of $NH_3$ at low pressure.
It is the ideal system to teach the quantum mechanics of two state systems.

But, experimentally it has been observed that, as pressure is raised, the
inversion line shifts and ultimately dissappears. The molecule localizes in
one of the classicaly stable configurations. This phenomenon has been
investigated for a long time \cite{Hund}.

A qualitative description using dipole-dipole interaction had been suggested
\cite{jonaclaverie}. It can be quantitatively explained \cite{jptprl}, using
the Keesom Energy \cite{Keesom}, if the interaction terms in a mean field
Hamiltonian cause the molecule to be stationary in one of the classical
localized configurations. A lot of work \cite{lpt0107041} \cite{gspreprint}
\cite{ath} has has gone into explaining the phenomenon. The role of collisions
in inducing molecular localization has also been suggested \cite{gspra}.
Perturbative terms have also been used to represent weak collisions
\cite{gsjpa}.

The purpose of this work is make the students, with basic knowledge of quantum
mechanics, familiar with this active branch of research. A semiclassical model
is presented for the molecules interacting via simple collisions in a Van der
Waals' Gas, hence introducing a mean field interaction constant, representing
hard ball collisions. The resulting Hamiltonian is used to find out the new
eigenstates as well as the inversion line frequency.

\section{Isolated molecule}
\label{isolated}

\subsection{Effective Potential}
\label{pot}

For an isolated molecule the one dimensional motion of the nucleus $N$ across
the plane containing the three $H$ nuclei can be considered separately from
the other degrees of freedom \cite{Feynman}.

The effective potential energy \cite{Cohen} is a function of the distance $x$
between the $N$ nucleus and the plane containing the $H$ nuclei. The symmetry
of the system about $x = 0$ dictates $V(x)$ to be an even function of $x$.
$V(x)$ has two minima corresponding to its classically stable configurations.
Let us assign this minimum value as zero.

The $N$ atom is repelled by the $H$ atoms if it approaches their plane. This
is represented by an energy maxima of height $V_1$ at $x = 0$. The chemical
bonding force ensures an increase in $V$ beyond the two minima. Hence, forming
a double well potential.

\subsection{Hilbert Space}
\label{space}

We say that the system is in state $\ket{L}$ when it is on one side and
$\ket{R}$ when it is on the other side.
\begin{eqnarray}
\ket{L}&=&\left(
\begin{array}{c}
1\\0
\end {array}
\right)\label{l}
\\
\ket{R}&=&\left(
\begin{array}{c}
0\\1
\end {array}
\right)\label{r}
\end{eqnarray}
Hence, forming the basis vectors of a two dimensional Hilbert space.

\subsection{Hamiltonian}
\label{ham}

We choose the Hamiltonian operator on the two-state Hilbert space to be
\begin{equation}
-\frac{\Delta E}{2} \sigma^x,
\label{Hamiltonian}
\end{equation} 
where $\sigma^x$ is the Pauli matrix
\begin{equation}
\sigma^x=
\left(
\bary{cc}
0&1\\
1&0\\
\ear
\right).\label{paulix}
\end{equation}

\subsection{Eigenstates}
\label{states}

The Eigenvectors corresponding to the Hamiltonian formulated in
Eq. (\ref{Hamiltonian}) are 
\begin{eqnarray}
\ket{1} &=& \frac{1}{\sqrt 2}\left(
\begin{array}{c}
 1\\ 1
\end{array}
\right)\label{1}
\\
\ket{2} &=& \frac{1}{\sqrt 2} \left(
\begin{array}{c}
1\\-1
\end{array}
\right),\label{2}
\end{eqnarray}
with Eigenvalues $-\Delta E/2$ and $+\Delta E/2$, respectively.

\subsection{Absorption Frequency}
\label{freq}

The inversion spectra of Ammonia is thus explained in terms of the energy
difference between these two stationary states.
\begin{equation}
\bar\nu = \frac{\Delta E}{h}
\label{nu}
\end{equation}

\section{Hard Ball Collisions}
\label{meanfield}

The gas is modelled as a system of nearly independent molecules, such that
each molecule is subjected to an external field representing random hard ball
collisions with the rest of the molecules. This interaction is to be
determined self-consistently. Then the linear response of this model to an
electromagnetic perturbation is to be studied to obtain the absorption
frequency. The inter-molecular interaction is found to control this frequency,
hence explaining the shift in the the inversion line.

The random motion of the molecules are faster than the inversion times, so on
the time scales of the inversion dynamics the molecules feel an effective
interaction arising out of the random collisions with the other molecules.

In the representation chosen for the Pauli matrices, the localizing effect of
the collision between two molecules $i$ and $j$ can be represented by an
interaction term of the form $\sigma^z_i \sigma^z_j$, where $\sigma^z$ has
localized Eigenstates
\begin{eqnarray}
\ket{L} = 
\left( \begin{array}{c} 1 \\ 0 \end {array} \right),\label{locall}
\\
\ket{R} = 
\left( \begin{array}{c} 0 \\ 1 \end {array} \right).\label{localr}
\end{eqnarray}

\section{Interaction Energy}
\label{interaction}

For extremely rarefied gases, the molecules are considered to be negligibly
small in size and the gas is supposed to have only kinetic energy and no
potential energy. The equation of state for 1 mole of such a gas is
\begin{equation}
PV = RT,
\label{ideal}
\end{equation}
where $P$ is the pressure and $V$ is the molar volume. But, the finite size of
the gas molecules effectively reduce the volume available for the free motion
of the molecules. If the molecules have a diameter $d$, no other molecular
center of mass can come within a sphere of radius $d$, centered at the center
of mass of the molecule under consideration. According to Van der Waals,
considering the molecules to be classical mechanical hard balls which only
apply contact forces and do not interact when away from each other, the
equation of state \cite{Saha} of the gas hence becomes
\begin{equation}
P(V-b) = RT,
\label{finite}
\end{equation}
where $b$ is defined as half the excluded volume, due to the finite size of
the molecules. Hence,
\begin{equation}
b = \frac{L}{2} \frac{4}{3} \pi d_1 d_2 d_3,
\label{b}
\end{equation}
where the $d_i$ are the principle collision diameters and $L$, the Avogadro's
number, is the number of molecules present in our system.

Expanding Eq. (\ref{finite}) we have
\begin{equation}
PV-Pb = RT,
\label{expandedfinite}
\end{equation}
Here, $Pb$ is the energy excluded from the equation of state due to hard ball
collisions, so, it must be accounted for. If $C$ be the mean excluded pressure
energy per molecule, then
\begin{equation}
LC = Pb.
\label{LC}
\end{equation}
Substituting, $b$ from Eq. (\ref{b}), we have
\begin{equation}
C = \frac{P}{2} \frac{4}{3} \pi d_1 d_2 d_3.
\label{Cinitial}
\end{equation}

Since, we are considering only one degree of freedom for the molecule,
$d_1=d_2=d$, while the other principle collision diameter $d_3$ is to be
determined self consistently from the state of the given molecule. Now, the
molecule has a maximum size for the localized Eigenstates of $\sigma^z$. So we
write,
\begin{equation}
d_3 = \ele{\psi}{\sigma^z}{\psi}d
\label{d3}
\end{equation}
Hence, the mean field interaction energy, per molecule, arising out of the
hard ball collisions, becomes
\begin{equation}
C = C_0 \ele{\psi}{\sigma^z}{\psi},
\label{C}
\end{equation}
where, the mean field interaction constant
\begin{equation}
C_0 = \frac{P}{2} \frac{4}{3} \pi d^3
\label{C0}
\end{equation}

\section{Mean Field Hamiltonian}
\label{meanfieldham}

In the mean-field approximation laid out in Eq (\ref{C}), we obtain the total
Hamiltonian
\begin{equation}
h(\psi) = -\frac{\Delta E}{2}\sigma^x-
C\sigma^z.
\label{meanfieldHamiltonianinitial}
\end{equation}
Substituting $C$ from Eq. (\ref{C0}), we have
\begin{equation}
h(\psi) = -\frac{\Delta E}{2}\sigma^x-
C_0\ele{\psi}{\sigma^z}{\psi}\sigma^z,
\label{meanfieldHamiltonian}
\end{equation}
where $\ket{\psi}$ is the state of the molecule being considered and $C_0$ is
given by Eq. (\ref{C0}). The state is to be determined self-consistently by
solving the nonlinear Eigenvalue problem associated with this Hamiltonian.

\section{Eigenstates}
\label{meanfieldstates}

The solution of the Eigenvalue problem associated with the Hamiltonian gives
the following results. 

If $C_0< \Delta E/2$, then
\begin{equation}
\ket{\lambda_1} = \ket{1},
\label{lambda1}
\end{equation}
the symmetric Eigenfunction, with Eigenvalue $\mu_1 = - \frac{\Delta E}{2}$,
is the only ground state. While,
\begin{equation}
\ket{\lambda_2} = \ket{2},
\label{lambda2}
\end{equation}
with Eigenvalue $\mu_2 = \frac{\Delta E}{2}$ forms the first excited state.

But, if $ C_0 \geq \Delta E/ 2 $, then there are two degenerate ground states,
\begin{equation}
\ket{\lambda_3^L} = \sqrt{\frac{1}{2} +\frac{\Delta E}{4C_0}}\ket{1}
+\sqrt{\frac{1}{2}-\frac{\Delta E}{4C_0}}~\ket{2}
\label{lambda3}
\end{equation}
and
\begin{equation}
\ket{\lambda_4^R} = \sqrt{\frac{1}{2} +\frac{\Delta E}{4C_0}}\ket{1}
-\sqrt{\frac{1}{2}-\frac{\Delta E}{4C_0}}\ket{2}
\label{lambda4}
\end{equation}
with $\mu_3 = \mu_4 = -C_0$.

Keeping this dependence on $C_0$ in mind, the critical value
\begin{equation}
C_0^\mathrm{cr} = \Delta E/2
\label{gcrit}
\end{equation}
is defined. This distinguishes the following two cases. For
$C_0<C_0^\mathrm{cr}$, the ground state of the system is the delocalized
symmetric state of an isolated molecule. For $C_0 \geq C_0^\mathrm{cr}$, there
are two different states which form the stationary states of the system.

These states described in Eqs. (\ref{lambda3}-\ref{lambda4}), transform into
one another under the action of the inversion operator $\sigma^x$,
\begin{equation}
\ket{\lambda_3^R} = \sigma^x\ket{\lambda_4^L}
\label{chiral}
\end{equation}
hence called chiral.

When $C_0 \gg C_0^\mathrm{cr}$, these new ground states become localized,
\begin{equation}
\lim_{\Delta E/C_0 \rightarrow 0}\ket{\lambda_3^L} = \ket{L}
\qquad
\lim_{\Delta E/C_0 \rightarrow 0}\ket{\lambda_4^R} = \ket{R}
\label{local}
\end{equation}

The emergence of the states described in Eqs. (\ref{lambda3}-\ref{lambda4})
imply a bifurcation of the ground state at a critical interaction
$C_0 = C_0^\mathrm{cr}$. Using the expression for $C_0$ from Eq. (\ref{C0}),
this bifurcation can be obtained above the critical pressure
\begin{equation}
P_\mathrm{cr} = \frac{3}{4 \pi} \frac{\Delta E}{d^3}.
\label{pcrit}
\end{equation}
Using $d = .274$ nm, $\Delta E$ corresponding to 23.8 GHz, we have
$P_\mathrm{cr} = 1.8$ atm.

\section{Inversion Line Frequency}
\label{meanfieldfreq}

Electro-magnetic radiation of angular frequency $\omega_0$ adds a perturbation
term \cite{jptprl} to the Hamiltonian described in
Eq. (\ref{meanfieldHamiltonian})
\begin{equation}
h_\mathrm{em}(t) = \epsilon f(t) \sigma^z
\label{emper}
\end{equation}
where $\epsilon$ is a small parameter and $f(t) = \theta(t) \cos(\omega_0 t)$,
$\theta(t)$ being the Heaviside function. The choice of a dipole coupling
approximation, $h_\mathrm{em} \propto \sigma^z$, is justified for a radiation
of wavelength long with respect to the molecular size.

Under the effect of the perturbation the isolated molecule state
$\ket{\lambda(t)}$ evolves according to the time-dependent nonlinear
Schr\"odinger equation
\begin{equation}
i\hbar\frac{\mathrm{d}\ket{\lambda(t)}}{\mathrm{d}t} =
\left[ h(\lambda(t)) + \epsilon f(t) \sigma^z \right]
\ket{\lambda(t)}
\label{pnls}
\end{equation}
where $h(\lambda)$ is the Hamiltonian for a molecule in the mean field
described in Eq. (\ref{meanfieldHamiltonian}).

The linear response to the perturbation is expressed by the generalized
susceptibility ${\mathcal R}(\omega) = \tilde{{\mathcal S}}_1(\omega)
/\tilde{f}(\omega)$, where $\tilde{f}(\omega)$ and $\tilde{{\mathcal
S}}_1(\omega)$ are the Fourier transforms of $f(t)$ and
${\mathcal S}_1(t)$, with ${\mathcal S}_1(t)$ defined by 
\begin{eqnarray}
{\mathcal S} (t) &\equiv&
\ele{\lambda(t)}{\sigma^z}{\lambda (t)}
\nonumber \\
&=& {\mathcal S}_0 (t)+ \epsilon~ {\mathcal S}_1 (t) + \ldots
\label{s}
\end{eqnarray}
Let us assume that at time $t = 0$ each molecule is in the delocalized ground
state  $\ket{\lambda_0} = \ket{1}$. The solution of the dynamical
Eq. (\ref{pnls}), with the initial condition $\ket{\lambda (0)} = \ket{1}$,
gives
\begin{equation}
\mathcal{R}(\omega) = 
\frac{2 \Delta E}
{(\hbar\omega)^2-
\left( \Delta E^2-2C_0\Delta E \right)}.
\label{susc}
\end{equation}
The generalized susceptibility has a unique pole at positive frequency which
corresponds to the inversion line frequency
\begin{equation}
\bar{\nu} = 
\frac{\Delta E}{h}\left( 1-\frac{2C_0}{\Delta E}\right)^{\frac12}.
\label{meanfieldnu}
\end{equation}
Substituting the value of $C_0$ from Eq. (\ref{C0}) this equation becomes
\begin{equation}
\bar\nu = \frac{\Delta E}{h}\sqrt{1-\frac{P}{P_\mathrm{cr}}}
= \bar\nu_0\sqrt{1-\frac{P}{P_\mathrm{cr}}},
\label{nufinal}
\end{equation}
which compares well with experimental data \cite{BL1} \cite{BL2}.

\section{Conclusions}
Students are customarily exposed to the symmetric and antisymmetric ground
states of a system. Quantum mechanics is taught as the basis of molecular
structure. Yet a large class of molecules are found to be chiral, not
symmetric. It is beyond the scope of undergraduate syllabii, as well as this
work, to address the quantum origins of chirality of organic molecules. But,
it can be instructive for students to learn about environment induced
localizations through the use of simple models such as this.

\begin{acknowledgements}
I would like to thank the Tata Institute of Fundamental Research (TIFR) and the Bhaba Atomic Research Centre (BARC) for organizing the National Initiative on Undergraduate Science (NIUS), the Homi Bhaba Centre for Science Education (HBCSE) for providing an unique learning environment at the 2nd Nurture Camp, my teammates Abhinandan Basu and Yashodhan Kanoria, my friend Shamashis Sengupta for clarifications about the Kinetic Theory of the Van der Walls' Gas, Mehuli Mondal for discussions on the Thermodynamics of such systems, and most of all, our mentors Prof. Arvind Kumar and Prof. Vikram Athaley for guiding us.
\end{acknowledgements}

\end{document}